\documentclass[aps,amssymb,amsmath,showpacs, superscriptaddress]{revtex4} 
\usepackage{latexsym}
\usepackage{amsmath}
\usepackage{amssymb}
\usepackage{amsfonts}
\usepackage{epsfig}
\usepackage{dcolumn}
\usepackage{bm}
\usepackage{epstopdf}

\begin{document}
\author{You-Chang Yang}
\email[]{youcyang@163.com}
\affiliation{Department of Physics, Nanjing University, Nanjing 210093, China}
\affiliation{School of Physics and Electronic Science, Zunyi Normal University, Zunyi 563006, China}

\author{Zhi-Yun Tan}
\email[]{tanzhiyun1979@126.com}
\affiliation{School of Physics and Electronic Science, Zunyi Normal University, Zunyi 563006, China}

\author{Hong-Shi Zong}
\email[]{zonghs@nju.edu.cn}
\affiliation{Department of Physics, Nanjing University, Nanjing 210093, China}

\author{Jialun Ping}
\email[]{jlping@njnu.edu.cn}
\affiliation{Department of Physics, Nanjing Normal University,Nanjing 210097, China}

\title{Dynamical study of $S$-wave $\bar{Q}Q\bar{q}q$ system}

\begin{abstract}
We perform the energy spectra calculating for the $S$-wave $\bar{Q}Q\bar{q}q$ (where $Q=c,b$ and $q=u,d,s$) system within two constituent quark models. The bound states of $B^*\bar{B}^*$ with $I(J^{PC})=1(0^{++}),~1(1^{+-}),~0(2^{++})$
and $B\bar{B}^*$ with isospin $I=1$, or $0$ are obtained in color-singlet-singlet channel. If considering the coupling of color channels, apart from the deep bound states appear in $[b\bar{q}]^{(*)}[q\bar{b}]^{(*)}$ scenario, a bound state
$[c\bar{q}]^*[q\bar{c}]^*$ with $I(J^{PC})=1(0^{++})$ is also formed. The $B\bar{B}^*$ and the $B^*\bar{B}^*$ with quantum number $1(1^{+-})$ are well candidates for $Z_b^{\pm}(10610)$ and $Z_b^{\pm}(10650)$ reported by Belle collaboration respectively, while the $B\bar{B}^*$ with isospin $0$ can be interpret as a candidate for  $Z_b^{0}(10610)$. A bound state $[c\bar{q}]^*[q\bar{c}]^*$ with $I(J^{PC})=1(0^{++})$ is comparable with  $Z_c^{+}(4025)$ announced by BES.
\end{abstract}
\maketitle

\section{Introduction}
In the past decades, more than twenty new charmonium- and bottomonium-like resonances,
usually called ``$XYZ$" states \cite {PDG,Zupanc,EPJC711534,Zb(10610),Zc3900BESIII,Zc3900Belle,
Zc3900Cleo-c,Zc4025BESIII,Z4430Belle,Z4050Belle},
have been reported by Belle, BaBar, BES, LHCb, CLEO and other collaborations. Many of $XYZ$ states,
such as $X(3872),~X(3915),~Y(4260),~Z^{+}(4430),~Z_c^{\pm}(3900)$, have been confirmed by different
experimental collaborations in same or different physical processes. Most of them do not seem to be understood
as conventional charmonium $c\bar{c}$ and bottomonium $b\bar{b}$ mesons predicted by quark model.
Especially, the states  $Z_c^{\pm}(3900)$ \cite{Zc3900BESIII,Zc3900Belle,Zc3900Cleo-c},
$Z_c^{\pm}(4025)$\cite{Zc4025BESIII}, $Z^{\pm}(4430)$ \cite{Z4430Belle},
$Z^{\pm}_1(4050),Z^{\pm}_2(4250)$\cite{Z4050Belle}, $Z_b^{\pm}(10610),
Z_b^{\pm}(10650)$\cite{Zb(10610)} have a nonzero electric charge, they must be genuine
exotic mesons with minimum four-quark content if they are really exist in nature. To understand
the structures of these new resonances, Many explanations, such as conventional quarkonium,
charmed hybrids, hadrocharmonium, diquarks and molecular pictures, have been proposed for these new
entities, but their nature still remains as a puzzle. The review papers \cite{PR429243,IJMPE17283,csb59,ijmpa30,fp10121,chxPhysrept6391,fbs571185,physrep6681,ppnp93143} and the references
therein reported the detail experimental data and the possible interpretations for $XYZ$ states.

According to the Quantum Chromodynamics (QCD), which is the underlying theory of strong interaction,
a colourless meson composed of $Q\bar{Q}$, $Qq\bar{Q}\bar{q}$ (where $Q=c,b$, and $q=u,d,s$) etc. are allowed.
Since the low energy process are completely governed by non-perturbation QCD effects, there is a lack of
any reliable approach to deal with the QCD non-perturbation problem. So it is still impossible for us to derive
the hadron spectrum analytically from the QCD at present. Lattice QCD was invented to solve QCD numerically through
simulations on the lattice, which has proven very powerful in the calculation of the hadron spectrum. Further
development is needed to obtain the satisfactory results for multiquark systems.
Besides, various theoretical frameworks and many phenomenological models with some kind of QCD spirit
were proposed to interpret these new $XYZ$ states, such as the QCD sum rule, chiral unitary model,
the one-boson-exchange (OBE) model, the one-pion-exchange (OPE) model, the diquark-antidiquark model,
and the chiral quark model (ChQM) etc.

In constituent quark model, Vijande {\em et al.} \cite{PRD76094022} studied the four-quark system
$c\bar{c}n\bar{n}$ by means of the hyperspherical harmonic formalism. However no bound states have been found
whether taking into account the exchange of scalar and pseudoscalar mesons or not.
Liu and Zhang \cite{PRC79035206} obtained a $D^{0}\bar{D}^{0*}$ bound state in a chiral SU(3) quark model with
including $\pi, \sigma$, $\omega$ and $\rho$ meson exchanges in it. In Ref. \cite{PRC80015208}, Liu and Zhang
systematically calculated the energy of four-quark system which composed of $S-$wave ($\bar{Q}q$) meson and
the ($\bar{q}Q$) meson. They obtained isoscalar $B\bar{B},~B^*\bar{B}^*(J=2)$, and $B\bar{B}^*(C=+)$ bound states.
In Ref. \cite{PRD81114025}, Yang and Ping obtained the bound state of $D^*\bar{D^*}$ with $J^{PC}=0^{++}$ and
$B^*\bar{B^*}$ with $J^{PC}=0^{++}$ and $2^{++}$ in the Bhaduri, Cohler, and Nogami
quark model by considering the coupling of color $1 \otimes 1$ and  $8 \otimes 8$
structure. In Ref. \cite{arxiv170309718} they studied possible neutral $D^{(*)}\bar{D}^{(*)}$ and
$B^{(*)}\bar{B}^{(*)}$ molecular states in the framework of the extended constituent quark models with the
$s$-channel one gluon exchange. The bound states of neutral $D^{(*)}\bar{D}^{(*)}$ with $J^{PC}=1^{++},2^{++}$
and $B^{(*)}\bar{B}^{(*)}$ with $J^{PC}=0^{++},1^{+-},1^{++},2^{++}$ are obtained when took into account of the
reasonable effective gluon mass.

By considering the multi-body confinement in constituent quark model, Deng, Ping, {\em et al.} believe the
$Y(2175)$, $f_0(600)$, $f_0(980)$ and $X(5176)$ are tetraquark states \cite{prd82074001}, and suggested the S-wave
tetraquark states $[cu][\bar{c}\bar{d}]$ with quantum numbers $IJ^P =11^+$ and $12^+$ are candidates for the
charged states $Z_c(3900)$ and $Z_c(4025)/Z_c(4020)$, respectively \cite{prd90054009}, and they believe
$Z_c^+(3900)$ or $Z_c^+(3885), Z_c^+(3930), Z_c^+(4020), Z_1^+(4050), Z_2^+(4250)$ and $Z_c^+(4200)$ can be
described as family of tetraquark $[cu][\bar{c}\bar{d}]$ states \cite{prd92034027}.

In $[\bar{Q}q][Q\bar{q}]$ system, since the substructure $[Q\bar{q}]$ (or $[q\bar{Q}]$) can form color-singlet
and color-octet alone, and the colorless $[\bar{Q}q][Q\bar{q}]$ can be formed by color-singlet-singlet($1 \otimes 1$)
or color-octet-octet ($8 \otimes 8$). Therefor the color channel coupling may be plays important role for the
energy spectra of $[\bar{Q}q][Q\bar{q}]$ system. In this work, we would like to study the energy spectra of the
$S$-wave $\bar{Q}Q\bar{q}q$ system in constituent quark models with considering the color channel coupling between
$1 \otimes 1$ and $8 \otimes 8$.

For comparison, we calculate the spectra of $[\bar{Q}q][\bar{q}Q]$ systems within two types of chiral
quark models \cite{ChQMI,slamanca} by means of Gaussian expansion method (GEM), which is a high accuracy method
for few-body systems developed by Kamimura, Hiyama {\em et al.} \cite{GEM} and extensively used in studying the
mass spectrum of multi-quark system \cite{ChQMI,prd80114023,prd81114025,jpg39045001,prd86014008,prd88074007}.
The two chiral quark models both include Goldstone-boson exchange in addition to color confinement and one-gluon-exchange
potential. The chiral partner, $\sigma$-meson, is also employed, although its existence is still in
controversy~\cite{sigma}. In the CQM I \cite{slamanca}, the $\sigma$-exchange
occurs between each pair of $u,d,s$, and the screening effect of color confinement is taken into account; while in the CQM II \cite{ChQMI}, the ordinary linear confinement is used and the $\sigma$
exchange occurs between the lightest quarks ($u$- or $d$-quark) only.

This paper is organized as follows. After the introduction, we present the chiral constituent quark models
in Sec. \ref{quarkmodel}. The wave functions of $[\bar{Q}q][\bar{q}Q]$ are constructed and the Gaussian expansion
method is introduced in Sec. \ref{wavefunction}. The spectra of the $[\bar{Q}q][\bar{q}Q]$ system are obtained
by solving the Schr\"{o}dinger equation and presented in Sec. \ref{results} with a brief discussion.
Finally, a summary is given in Sec. \ref{summary}.

\section{the constituent quark model and parameters}\label{quarkmodel}
The constituent quark model (CQM) was proposed for obtaining a simultaneous description of the baryon spectra and
the nucleon-nucleon interaction \cite{prc64058201}.  This model has been generalized to all flavor sectors giving
a reasonable description of the meson spectra \cite{slamanca}, the baryon spectra \cite{prc63035207}. It is also
applied to scalar mesons with four-quark configurations included \cite{prd72034025}, and some possible four-quark system
spectra \cite{prd73054004,prd79074010}. This model contains one-gluon-exchange (OGE) potential governed by
the QCD perturbation effects, the Goldstone boson-exchange potentials for the spontaneous breaking of the original
$SU(3)_L \otimes SU(3)_R$ chiral symmetry at some momentum scale, and a screened confined interaction as dictated
by unquenched lattice calculations \cite{prep3431}. The detailed Hamiltonian of this model has the form
\begin{equation}
H=\sum_{i=1}^4 \left( m_i+\frac{\mathbf{p}_i^2}{2m_i}
\right)-T_{CM} +\sum_{j>i=1}^4
(V_{ij}^C+V_{ij}^G+V_{ij}^\chi+V_{ij}^\sigma),
\label{h_cqm}
\end{equation}
where $\chi=\pi,K,\eta$, and $T_{CM}$ is the kinetic energy operator of the center-of-mass motion.

The screened confinement interaction in this model is
\begin{equation}
V_{ij}^{C} ={\boldsymbol{\lambda}}^c_{i} \cdot
{\boldsymbol{\lambda}}^c_{j}~\{-a_{c}(1-e^{-\mu_c r_{ij}})+
\Delta \},~\label{confinement2}
\end{equation}
where $\mu_c$ is a color screening parameter.

The potential of the OGE part reads
\begin{eqnarray}
V_{ij}^G =\alpha_{s} \frac{{\boldsymbol{\lambda}}^c_{i} \cdot
{\boldsymbol{\lambda}}^c_{j}}{4} \left[{\frac{1}{r_{ij}}}-
{\frac{2\pi}{3m_im_j}}~({\boldsymbol{\sigma}}_{i}\cdot
{\boldsymbol{\sigma}}_{j})~\delta(\mathbf{r}_{ij})
 \right],  \label{gluon}
\end{eqnarray}
where, $\boldsymbol{\sigma}$, $\boldsymbol{\lambda}$ are the
SU(2) Pauli matrices and the SU(3) Gell-Mann matrices, respectively.
The $\boldsymbol{\lambda}$ should be replaced by
$-\boldsymbol{\lambda}^*$ for the antiquark. The delta function
$\delta(\mathbf{r}_{ij})$ should be regularized \cite{prl441369}. The regularization is justified
based on the finite size of the constituent quark and should
therefore be flavor dependent [~\cite{slamanca,weinstein}
\begin{equation}
\delta(\mathbf{r}_{ij})=\frac{1}{4\pi r_{ij}~r_0^2(\mu)}~
e^{-r_{ij}/r_0(\mu)},
\end{equation}
where $r_0(\mu)=r_0/\mu$ and $\mu$ is the reduced mass of
quark-quark (or antiquark) system.

In the non-relativistic quark model, the wide energy covered from light to heavy quark requires an
effective scale-dependent strong coupling constant $\alpha_s$ in
Eq. (\ref{gluon}) that cannot be obtained from the usual one-loop
expression of the running coupling constant because it diverges when
$Q\rightarrow\Lambda_{QCD}$. An effective scale-dependent
strong coupling constant \cite{slamanca} is taken as
\begin{equation}
\alpha_s(\mu)=\frac{\alpha_0}{\ln\left[\frac{\mu^{2}+\mu_0^2}{\Lambda_0^2}\right]}~,
\end{equation}
where $\mu_0$ and $\Lambda_0$ are the free parameters.

The meson-exchange potentials have the form
\begin{equation}
V_{ij}^{\pi}
=C(g_{ch},\Lambda_{\pi},m_{\pi}){\frac{m_{\pi}^{2}}{{12m_{i}m_{j}}}}
H_1(m_{\pi},\Lambda_{\pi},r_{ij})
({\boldsymbol{\sigma}}_{i}\cdot
{\boldsymbol{\sigma}}_{j})~\sum_{a=1}^3\lambda_{i}^{a}\cdot
\lambda_{j}^{a}, \label{pi}
\end{equation}
\begin{equation}
V_{ij}^{K}=C(g_{ch},\Lambda_{K},m_{K}){\frac{m_K^{2}}{{\
12m_{i}m_{j}}}} H_1(m_K,\Lambda_K,r_{ij})
({\boldsymbol{\sigma}}_{i}\cdot
{\boldsymbol{\sigma}}_{j})~\sum_{a=4}^7\lambda_{i}^{a}\cdot
\lambda_{j}^{a}, \label{k}
\end{equation}
\begin{eqnarray}
V_{ij}^{\eta}
=&&C(g_{ch},\Lambda_{\eta},m_{\eta}){\frac{m_{\eta}^{2}}{{
12m_{i}m_{j}}}} H_1(m_{\eta},\Lambda_{\eta},r_{ij})
({\boldsymbol{\sigma}}_{i}\cdot
{\boldsymbol{\sigma}}_{j})~\nonumber\\&&
\times\left[\cos\theta_P(\lambda_{i}^{8}\cdot
\lambda_{j}^{8})-\sin\theta_P(\lambda_{i}^{0}\cdot \lambda_{j}^{0})
\right],\label{eta}
\end{eqnarray}
\begin{equation}
V_{ij}^{\sigma} = -C(g_{ch},\Lambda_{\sigma},m_{\sigma})~
H_2(m_{\sigma},\Lambda_{\sigma},r_{ij}) \label{sigma}
\end{equation}
\begin{equation}
H_1(m,\Lambda,r)=\left[ Y(mr)-{\frac{\Lambda^{3}}{m^{3}}} Y(\Lambda
r)\right]
\end{equation}
\begin{equation}
H_2(m,\Lambda,r)=\left[ Y(mr)-{\frac{\Lambda}{m}} Y(\Lambda
r)\right]
\end{equation}
\begin{equation}
C(g_{ch},\Lambda,m)={\frac{g_{ch}^{2}}{{4\pi
}}}{\frac{\Lambda^{2}}{{\Lambda^{2}-m^{2}}}} m
\end{equation}
where adopt $\lambda^{0}=\sqrt{\frac{2}{3}}I$ due to the normalization of
$SU(3)$ matrix. $Y(x)$ is the standard Yukawa function defined by
$Y(x)=e^{-x}/x$ and $\Lambda$ is a cutoff parameter. The
chiral coupling constant $g_{ch}$ is determined from the $\pi NN$
coupling constant through
\begin{equation}
\frac{g_{ch}^{2}}{4\pi }=\left( \frac{3}{5}\right) ^{2}{\frac{g_{\pi NN}^{2}%
}{{4\pi }}}{\frac{m_{u,d}^{2}}{m_{N}^{2}}},
\end{equation}
and flavor $SU(3)$ symmetry is assumed.

In this version of constituent quark model (CQM I), The $\sigma$-meson is exchangeable among
$u,d,s$ quarks,
\begin{equation}
V_{ij}^{\chi, \sigma}=\left\{
\begin{array}{ll}
ud,dd,uu\Rightarrow V_{\pi}+V_{\sigma}+V_{\eta } &  \\
us,ds\Rightarrow V_{\sigma}+V_{K}+V_{\eta } &  \\
ss\Rightarrow V_{\sigma}+V_{\eta } &
\end{array}
\right.  \label{pot}
\end{equation}

Another version of the constituent quark model (CQM II) \cite{ChQMI} is employed here
for calculating the spectra of possible four-quark state. The Hamiltonian is similar to the
aforementioned CQM I. However, the $\sigma$ meson-exchange only occurs between the lightest quarks
($u$- or $d$-quark) due to its non-strange nature, and the simple linear confining potential is taken in it.
\begin{equation}
V_{ij}^{C} ={\boldsymbol{\lambda}}^c_{i} \cdot
{\boldsymbol{\lambda}}^c_{j}~(-a_{c}r_{ij}-
\Delta).~\label{confinement}
\end{equation}

All parameters of the CQM I and CQM II listed in TABLE \ref{parameters} are taken from the Ref.\cite{slamanca} and Ref.\cite{ChQMI}, respectively.
\begin{center}
\begin{ruledtabular}
\begin{table}[htp]
\caption{Parameters of the CQM I and CQM II. Each parameter of meson-exchange of two quark models has same value, namely, $m_\pi=0.7$~fm$^{-1}$, $m_{\eta}=2.77$~fm$^{-1}$,$m_{\sigma}=3.42$~fm$^{-1}$,
$\Lambda_{\pi}=\Lambda_{\sigma}=4.2$~fm$^{-1}$, $\Lambda_{\eta}=5.2$~fm$^{-1}$, $\theta_p=-15^o$, $g^2_{ch}/4\pi$=0.54. \label{parameters}}
{\begin{tabular}{@{}c cccc ccc@{}}
  & \multicolumn{4}{c}{Quark masses} & \multicolumn{3}{c}{Confinement} \\ \hline
  & $m_{u,d}$ & $m_s$ & $m_c$ &  $m_b$ & $a_c$  & $ \Delta$ & $\mu_c$ \\
  & (MeV) & (MeV) & (MeV) &  (MeV) & (MeV fm$^{-1}$)  & (MeV) & (fm$^{-1}$) \\ \hline
  CQM I & 313 & 555 & 1752 & 5100 & 430 & 181.1 & 0.7 \\
  CQM II  & 313 & 525 & 1731 & 5100 & 160 & $-131.1$ & --- \\ \hline
   & \multicolumn{4}{c}{OGE} & &  & \\  \hline
  & $\alpha_{0}$ & $r_{0}$ & $\mu_0$  & $\Lambda_0$ & &  & \\
  &  & (MeV fm) & (MeV)  & (fm) & &  & \\  \hline
   CQM I & 2.118 & 28.17 & 36.976 & 0.113 \\
   CQM II  & 2.65 & 28.17 & 36.976 & 0.075 & & &
\end{tabular}}
\end{table}
\end{ruledtabular}
\end{center}

\section{wave function of $[Q\bar{q}][\bar{Q}q]$ system}\label{wavefunction}
The total wave function of $\bar{Q}Q\bar{q}q$ system can be written as a sum of outer products of
color, isospin, spin, and spacial terms
\begin{equation}
\Psi^{I,I_z}_{J,J_z}=\left|\xi\right\rangle
\left|\eta\right\rangle^{II_z}\Phi_{JJ_z},\label{twave}
\end{equation}
with
\[\Phi_{JJ_z}=\left[\left|\chi\right\rangle_{S}\otimes\left|\Phi\right\rangle_{L_T}\right]_{JJ_z}
\]
where $\left|\xi\right\rangle$, $\left|\eta\right\rangle^{I}$, $\left|\chi\right\rangle_{S}$, $\left|\Phi\right\rangle_{L_{T}}$ denote color (color singlet), isospin ($I$), spin ($S$) and
spacial (angular momentum $L_T$) wave functions, respectively.

\begin{figure}[ht]
\includegraphics[scale=0.4]{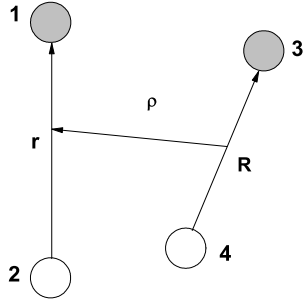}
\caption{The relative coordinate for
$[\bar{Q}q][\bar{q}Q]$ system. Darkened and open circles represent
 antiquarks and quarks, respectively.}\label{coordinate}
\end{figure}

The $H$-type Jacobi coordinates shown in FIG. \ref{coordinate} is chosen here, and we assume that particles 1, 3 are the $\bar{Q}$, $\bar{q}$ quark, and particles 2, 4 are the $q$, $Q$ quark, respectively. The Jacobi coordinates are defined

\begin{eqnarray}
&&\mathbf{r}=\mathbf{r}_1-\mathbf{r}_2,\ ~~~~
\mathbf{R}=\mathbf{r}_3-\mathbf{r}_4,~~~
\boldsymbol{\rho}=\frac{m_1\mathbf{r}_1+m_2\mathbf{r}_2}{m_1+m_2}
-\frac{m_3\mathbf{r}_3+m_4\mathbf{r}_4}{m_3+m_4},\\&&\mathbf{R}_{cm}=\sum_{i=1}^4
m_i \mathbf{r}_i/\sum_{i=1}^4 m_i,
\end{eqnarray}
where $m_i$ is the mass of the $i$th quark, and $\mathbf{R}_{cm}$ is the coordinate of the mass-center.
The outer products of spacial and spin wave function for a four-quark state with coupling
$[\bar{Q}q]$ and  $[\bar{q}Q]$ shown in FIG.\ref{coordinate} is
\begin{eqnarray}
\Phi_{JJ_z}=[[
\phi^G_{lm}(\mathbf{r})\chi_{s_1m_{s_1}}]_{J_1M_1}[\psi^G_{LM}(\mathbf{R})\chi_{s_2m_{s_2}}]_{J_2M_2}
]_{J_{12}M_{12}}\varphi^G_{\beta\gamma}(\boldsymbol{\rho})]_{JJ_z}.
\end{eqnarray}

In GEM \cite{GEM}, the spacial wave functions of three relative motion in FIG.\ref{coordinate}
are expanded through Gaussian wave-functions with various size
\begin{eqnarray}
\phi^G_{lm}(\mathbf{r})=\sum_{n=1}^{n_{max}}c_nN_{nl}r^le^{-\nu_nr^2}Y_{lm}(\hat{\mathbf{r}})\label{gem1}
\end{eqnarray}
\begin{eqnarray}
\psi^G_{LM}(\mathbf{R})=\sum_{N=1}^{N_{max}}c_NN_{NL}R^Le^{-\zeta_NR^2}Y_{LM}(\hat{\mathbf{R}})\label{gem2}
\end{eqnarray}
\begin{eqnarray}
\varphi^G_{\beta\gamma}(\boldsymbol{\rho})=
\sum_{\alpha=1}^{\alpha_{max}}c_{\alpha}N_{\alpha\beta}\rho^{\beta}e^{-\omega_\alpha
\rho^2}Y_{\beta\gamma}(\hat{\mathbf{\rho}})\label{gem3}
\end{eqnarray}
where the expansion coefficients $c_n, c_N, c_{\alpha}$ are determined by solving the
Schr\"{o}dinger equation, and $N_{nl},~N_{NL},~N_{\alpha\beta}$ are respectively normalization
constant for wave function $\phi^G_{lm}(\mathbf{r}),~ \psi^G_{LM}(\mathbf{R}),~
\varphi^G_{\beta\gamma}(\boldsymbol{\rho})$, which reads
\begin{equation}
N_{nl}=\left(\frac{2^{l+2}(2\nu_n)^{l+\frac{3}{2}}}{\sqrt{\pi}(2l+1)!!}\right)^{\frac{1}{2}}
\end{equation}

The Gaussian size parameters take geometric progression
\begin{eqnarray}
\nu_{n}=\frac{1}{s^2_n},& s_n=s_1a^{n-1},&
a=\left(\frac{s_{n_{max}}}{s_1}\right)^{\frac{1}{n_{max}-1}}.
\label{geo progress}
\end{eqnarray}
The expression of $\zeta_N,~\omega_\alpha$ in Eqs.
(\ref{gem2}) - (\ref{gem3}) are similar to Eq. (\ref{geo progress}).

The flavor wave functions of four-quark $[\bar{Q}q][\bar{q}Q]$ ($Q=c,b;~~q=u,d,s$) system,
which were studied by Liu, Luo and Zhu \cite{epjc61411}, are shown in TABLE \ref{DD} and \ref{BB}.
According to the Ref. \cite{epjc61411}, the pseudoscalar-pseudoscalar type $(P\bar{P})$ is categorized
as two systems, namely, $\mathcal{D}\bar{\mathcal{D}}$ and $\mathcal{B}\bar{\mathcal{B}}$. Where the
$\mathcal{D}$, $\bar{\mathcal{D}}$, $\mathcal{B}$, $\bar{\mathcal{B}}$ represent $(D^{+},D^0,D_s^0)$,
$(D^{-},\bar{D}^0, \bar{D}_s^0)$, $(B^{+},B^0,B_s^0)$ and  $(B^{-},\bar B^0,\bar B_s^0)$ triplet
respectively. The vector-vector type $(V\bar{V})$ has the same expression as that of the $P\bar{P}$
with the substitutions $D\rightarrow D^{*}$, $\bar{D}\rightarrow \bar{D}^{*}$, $B\rightarrow B^{*}$
and $\bar{B}\rightarrow \bar{B}^{*}$. The wave functions with definite $C$ parity are given through
a linear combination \cite{zpc61523} for the neutral states. The normal convention of PDG~\cite{PDG}
e.g. $B^0=d\bar{b}$ and $\bar{B}^0=\bar{d}b$, are used in the Table.
\begin{table}[htp]
\begin{center}
\begin{ruledtabular}
\caption{The flavor wave functions of the
$\mathcal{D}\bar{\mathcal{D}}$, $\mathcal{D}\bar{\mathcal{D}}^*$
systems. \label{DD}}   
{\begin{tabular}{@{}lcc@{}}
Isospin           & $\mathcal{D}\bar{\mathcal{D}}~(P\bar{P}~ or~ V\bar{V})$ &
$\mathcal{D}\bar{\mathcal{D}}^* ~(P\bar{V})$ \\\hline
                  & $D^+\bar{D}_{s}^0$ & $D^{*+}\bar{D}_{s}^0,~D^{+}\bar{D}_{s}^{*0}$ \\
I=$\frac{1}{2}$   & ${D}^0 \bar{D}_s^0$ & $D^{*0}\bar{D}_{s}^0,~D^{0}\bar{D}_{s}^{*0}$ \\
                  & ${D}_s^0 \bar{D}^0$ & ${D}_s^{*0} \bar{D}^0,~{D}_s^{0} \bar{D}^{*0}$ \\
                  & ${D}_s^0 D^-$ & ${D}_s^{*0} D^-,~{D}_s^0 D^{*-}$ \\
                  & $D^+\bar{D}^0$ & $D^{*+}\bar{D}^0,~ D^{+}\bar{D}^{*0}$ \\
I=1               & $\frac{1}{\sqrt{2}}(D^+D^--D^0\bar{D}^0)$ &
                    $\frac{1}{2}[(D^{*+}D^--D^{*0}\bar{D}^0)+c\,(D^{*-}D^+ -\bar{D}^{*0}D^0)]$, \\
                  & & $\frac{1}{2}[(D^+D^{*-}-D^0\bar{D}^{*0})+c\,(D^-D^{*+}-\bar{D}^0 D^{*0})]$ \\
                  & ${D}^0 D^-$ & ${D}^{*0} D^-,~ {D}^0 D^{*-}$ \\
I=0 ($l$)          & $\frac{1}{\sqrt{2}}(D^+D^-+D^0\bar{D}^0)$ &
                    $\frac{1}{2}[(D^{*+}D^-+D^{*0}\bar{D}^0)+c\,(D^{*-}D^+ +\bar{D}^{*0}D^0)]$,\\
                  & & $\frac{1}{2}[(D^+D^{*-}+D^0\bar{D}^{*0})+c\,(D^-D^{*+}+\bar{D}^0 D^{*0})]$ \\
I=0 ($s$)          & $D_s^0 \bar{D}_s^0$ &
$\frac{1}{\sqrt{2}}(D_s^{*0} \bar{D}_s^0+c\, \bar{D}_s^{*0}
               D_s^0),~\frac{1}{\sqrt{2}}(D_s^0 \bar{D}_s^{*0}+\bar{D}_s^0 D_s^{*0})$
\end{tabular}}
\end{ruledtabular}
\end{center}
\end{table}

\begin{table}[htp]
\begin{center}
\begin{ruledtabular}
\caption{The same as TABLE \ref{DD} for $\mathcal{B}\bar{\mathcal{B}}$, $\mathcal{B}\bar{\mathcal{B}}^*$
systems. \label{BB} }
{\begin{tabular}{@{}lcc@{}}
Isospin           & $\mathcal{B}\bar{\mathcal{B}}~ (P\bar{P}~ or~ V\bar{V})$ &
$\mathcal{B}\bar{\mathcal{B}}^* ~(P\bar{V})$ \\\hline
                  & $B^+\bar{B}_{s}^0$ & $B^{*+}\bar{B}_{s}^0,~B^{+}\bar{B}_{s}^{*0}$ \\
I=$\frac{1}{2}$   & ${B}^0 \bar{B}_s^0$ & $B^{*0}\bar{B}_{s}^0,~B^{0}\bar{B}_{s}^{*0}$ \\
                  & ${B}_s^0 \bar{B}^0$ & ${B}_s^{*0} \bar{B}^0,~{B}_s^{0} \bar{B}^{*0}$ \\
                  & ${B}_s^0 B^-$ & ${B}_s^{*0} B^-,~{B}_s^0 B^{*-}$ \\
                  & $B^+\bar{B}^0$ & $B^{*+}\bar{B}^0,~ B^{+}\bar{B}^{*0}$ \\
I=1               & $\frac{1}{\sqrt{2}}(B^+B^--B^0\bar{B}^0)$ &
                    $\frac{1}{2}[(B^{*+}B^--B^{*0}\bar{B}^0)+c\,(B^{*-}B^+ -\bar{B}^{*0}B^0)]$, \\
                  & & $\frac{1}{2}[(B^+B^{*-}-B^0\bar{B}^{*0})+c\,(B^-B^{*+}-\bar{B}^0 B^{*0})]$ \\
                  & ${B}^0 B^-$ & ${B}^{*0} B^-,~ {B}^0 B^{*-}$ \\
I=0 ($l$)          & $\frac{1}{\sqrt{2}}(B^+B^-+B^0\bar{B}^0)$ &
                    $\frac{1}{2}[(B^{*+}B^-+B^{*0}\bar B^0)+c\,(B^{*-}B^+ +\bar{B}^{*0}B^0)]$,\\
                  & & $\frac{1}{2}[(B^+B^{*-}+B^0\bar{B}^{*0})+c\,(B^-B^{*+}+\bar{B}^0 B^{*0})]$ \\
I=0 ($s$)          & $B_s^0 \bar{B}_s^0$ &
$\frac{1}{\sqrt{2}}(B_s^{*0} \bar{B}_s^0+c\, \bar{B}_s^{*0}
               B_s^0),~\frac{1}{\sqrt{2}}(B_s^0 \bar{B}_s^{*0}+\bar{B}_s^0 B_s^{*0})$
\end{tabular}}
\end{ruledtabular}
\end{center}
\end{table}

The spin wavefunction can be easily constructed by angular momentum coupling, which are omitted here.

Taking $[D^*\bar{D}^*]_{1}$ as an example to illustrate the spin and flavor wavefunctions, we have
\begin{eqnarray}
|\chi \rangle_{11} & = & \sqrt{\frac{1}{2}}(|11\rangle |10\rangle-|10\rangle |11\rangle) \nonumber \\
  & = & \frac{1}{2}(\alpha\alpha\alpha\beta+\alpha\alpha\beta\alpha
  -\alpha\beta\alpha\alpha-\beta\alpha\alpha\alpha) \nonumber \\
|\eta \rangle_{00} & = & \sqrt{\frac{1}{2}}(\bar{D}^0D^0+D^-D^+)
   =\sqrt{\frac{1}{2}}(\bar{c}u\bar{u}c+\bar{c}d\bar{d}c).  \nonumber
\end{eqnarray}

The color wave function of color-singlet states for a four-quark state in configuration of
FIG.\ref{coordinate} can be constructed in two ways, they are
 \begin{eqnarray}
 \left|\xi_1\right\rangle &=&\left|\mathbf{1}_{12}\otimes\mathbf{1}_{34}\right\rangle =\frac{1}{3}\left(\left|\bar{r}r\bar{r}r\right\rangle+\left|\bar{g}g\bar{g}g\right\rangle+\left|\bar{b}b\bar{b}b\right\rangle
+\left|\bar{r}r\bar{g}g\right\rangle+\left|\bar{r}r\bar{b}b\right\rangle+\left|\bar{g}g\bar{r}r\right\rangle+\left|\bar{g}g\bar{b}b\right\rangle+
\left|\bar{b}b\bar{r}r\right\rangle+\left|\bar{b}b\bar{g}g\right\rangle\right)\label{color1x1}\\
 \left|\xi_2\right\rangle &=& \left|\mathbf{8}_{12}\otimes\mathbf{8}_{34}\right\rangle=\frac{1}{6\sqrt{2}} \left[ 3(\bar{b}r\bar{r}b+\bar{b}g\bar{g}b+\bar{g}r\bar{r}g+\bar{g}b\bar{b}g+\bar{r}g\bar{g}r+\bar{r}b\bar{b}r)
 +2(\bar{r}r\bar{r}r+\bar{g}g\bar{g}g+\bar{b}b\bar{b}b)\right . \nonumber\\
 && ~~~~~~~~~~~~~~~~~~~~~~~~~~\left . -(\bar{r}r\bar{g}g+\bar{g}g\bar{r}r+\bar{b}b\bar{r}r+\bar{b}b\bar{g}g+\bar{r}r\bar{b}b+\bar{g}g\bar{b}b)\right ]\label{color8x8}
 \end{eqnarray}
The color-singlet singlet channel as a molecular state picture has been generally studied in hadronic state level.
However, the $[\bar{Q}q]$  and $[\bar{q}Q]$ are allowed in color singlet and octet in QCD theory,
so the coupling effect for color-singlet singlet and color-octet octet should be taken into account
in the constituent quark model. The color matrix elements of $\langle \xi_k|\mathbf{\lambda}_i^c\cdot\mathbf{\lambda}^c_j|\xi_l\rangle$
with $j>i=1,2,3,4$ and $k,l=1,2$, are given in Table \ref{colormatrix}.
\begin{center}
\begin{ruledtabular}
\begin{table}[htb]
\caption{Color matrix elements $\langle \xi_k|\mathbf{\lambda}_i^c\cdot\mathbf{\lambda}^c_j|\xi_l\rangle$
with $j>i=1,2,3,4$. Here $|\xi_k\rangle,~ |\xi_l\rangle$ with $k,l=1,2$ represent the color wave
functions defined in Eqs. (\ref{color1x1}-\ref{color8x8}). }\label{colormatrix}
\begin{center}
\begin{tabular}{c|cccccc}
$(i,j)$ &(1,2)&(3,4) &(1,3)&(2,4)&(1,4)&(2,3) \\ \hline
 $(k,l)=(1,1)$ & $-\frac{16}{3}$&$-\frac{16}{3}$ &0&0&0&0 \\
 $(k,l)=(2,2)$ & $\frac{2}{3}$&$\frac{2}{3}$
   &$-\frac{4}{3}$&$-\frac{4}{3}$&$-\frac{14}{3}$&$-\frac{14}{3}$\\
 $(k,l)=(1,2)$ & 0&0 &$\sqrt{\frac{32}{9}}$&$\sqrt{\frac{32}{9}}$
 &$-\sqrt{\frac{32}{9}}$&$-\sqrt{\frac{32}{9}}$
 \\
\end{tabular}
\end{center}
\end{table}
\end{ruledtabular}
\end{center}

\section{Results and discussions}\label{results}
Unless the mass of a four-quark state $[\bar{Q}q][\bar{q}Q]$ below the threshold of mesons
composed of $[\bar{Q}q]$  and $[Q\bar{q}]$, or it is unstable, since it can easily fall-apart
decays into two mesons with appropriate quantum numbers. So one believes that a good fit of
meson mass spectra, with the same parameters used in four-quark calculations, must be the most
important criterium\cite{weinstein,Manohar2,Bhduri2, Brink, Janc1}. So using the aforementioned
parameters, we firstly calculate the  Schr\"{o}dinger equation
\begin{eqnarray}
\left(H-E\right)\Psi^{I,I_z}_{J,J_z}=0 \label{schrodinger}
\end{eqnarray}
with Rayleigh-Ritz variational principle to get spectra of the conventional mesons.
We obtain the converged results, which are listed in TABLE \ref{meson spectrum} and marked
as "CQM II" and "CQM I", with the number of gaussians $n_{max}=7$, and the size parameter
$s_n$ running from $0.1$ to $2$~fm in spatial wavefunction of Eq.(\ref{gem1}). For comparing our
results with previous work performed by Vijande, {\em et al.} in Ref.\cite{slamanca}
with chiral quark model, their results are also listed in the table.
Obversely, the results of "CQM I" are well consistent with that of Ref.\cite{slamanca}.
\begin{center}
\begin{ruledtabular}
\begin{table}[htp]
\caption{Numerical results of conventional meson spectra (unit: MeV) in two quark models.
The experimental data takes from the latest Particle Data Group\cite{PDG}.\label{meson spectrum}}
{\begin{tabular}{@{}lccccllcccccl@{}}
 Meson          &\multicolumn{2}{c}{CQM II}        &Ref.\cite{slamanca} & \multicolumn{2}{c}{CQM I} & Exp. \\ \hline
                &Mass   &$\sqrt{<r^2>}$ &Mass &Mass &$\sqrt{<r^2>}$& & \\ \hline
 $\pi$          &140.1 &0.67           &139      &153.2 &0.86 &139.57$\pm$0.00035 \\
 K              &496.2 &0.67           &496      &484.9 &0.86 &493.677$\pm$0.016  \\
 $\rho(770)$    &775.3 &0.88           &772      &773.1 &0.77 &775.49$\pm$0.34    \\
 $K^*(892)$     &917.9 &0.84           &910      &907.7 &0.79 &896.00$\pm$0.25   \\
 $\omega(782)$  &703.7 &0.84           &691      &696.6 &0.79 &782.65$\pm$0.12   \\
 $\phi(1020)$   &1016.8&0.79           &1020      &1011.9 &0.82 &1019.422$\pm$0.02 \\
 $\eta_c(1s)$   &2995.7&0.56           &2990      &2999.8 &0.9 &2980.3$\pm$1.2     \\
 $J/\psi(1s)$   &3097.6&0.61           &3097      &3096.7 &0.88 &3096.916$\pm$0.011 \\
 $D^0$          &1882.2&0.72           &1883      &1898.4 &0.84 &1864.84$\pm$0.17  \\
 $D^*$          &2000.2&0.78           &2010      &2017.3 &0.81 &2006.97$\pm$0.19  \\
 $D_s$          &1966.6 &0.66          &1981      &1991.8&0.86 &1968.49$\pm$0.34 \\
 $D^*_s$        &2091.1 &0.72          &2112      &2115.7&0.84  &2112.3$\pm$0.5\\
 $B^\pm$        &5284.7 &0.74          &5281      &5277.9&0.83  &5279.15$\pm$0.31\\
 $B^0$          &5284.7 &0.74          &5281      &5277.9&0.83  &5279.53$\pm$0.33\\
 $B^*$          &5324.3 &0.76          &5321      &5318.8&0.82  &5325.1$\pm$0.5\\
 $B^0_s$        &5360.6 &0.67          &5355      &5355.8&0.86  &5366.3$\pm$0.6\\
 $B^*_s$        &5403.6 &0.69          &5400      &5400.5&0.87  &5412.8$\pm$1.3\\
 $\eta_b(1s)$   &9384.6 &0.42          &9454      &9467.9&0.93  &9388.9$^{+3.1}_{-2.3}$(stat)\\
 $\Upsilon(1s)$ &9462.4 &0.45          &9505      &9504.7&0.92  &9460.30$\pm$0.26
\end{tabular}}
\end{table}
\end{ruledtabular}
\end{center}

In order to judge a four-quark configuration is a stable bound state or not, the binding energy of a four-quark system is  defined
\begin{equation}
\Delta E=E_T-E_{th}£¬ \label{deltaE}
\end{equation}
where $E_T$ stands for the four-quark energy and $E_{th}$ for the energy of the corresponding threshold.
If $\Delta E < 0$, then this four-quark state is stable against the strong interaction, i.e, one has a
proper bound state. However, if $\Delta E \geq 0$, it indicates that the four-quark solution correspond
to an unbound four-quark configuration. The thresholds related to $[\bar{Q}q][\bar{q}Q]$ system are
obtained from TABLE \ref{meson spectrum} and listed in TABLE \ref{threshold}.
\begin{center}
\begin{ruledtabular}
\begin{table}[htp]
\caption{Threshold energy of  $\mathcal{D}\bar{\mathcal{D}},
\mathcal{D}\bar{\mathcal{ D}}^*,\mathcal{D^*}\bar{\mathcal{D}}^*$,$\mathcal{B}\bar{\mathcal{B}},
\mathcal{B}\bar{\mathcal{ B}}^*$ and~ $\mathcal{B^*}\bar{\mathcal{B}}^*$ in
two quark models (unit: MeV). \label{threshold}}
{\begin{tabular}{@{}cccccc@{}}
~~~J~~~ &~~~I~~~ & $M_1M_2$ & E$_{th}$(CQM II) & E$_{th}$(CQM I)&E$_{th}$(Exp.)\\
\hline
&0,1             &$\bar{D}^0D^+$                  &3764.8   &3796.8  &3729.68\\
0&0              &$D_s^-D_s^+$                    &3933.2   &3983.6  &3936.98\\
&$\frac{1}{2}$   &$D_s^-D^0/D_s^+D^-$             &3848.8   &3890.2  &3833.33\\
\hline
&0,1            &$\bar{D}^0D^{*+}$                &3882.4    &3915.7 &3871.81\\
1&0             &$D_s^-D_s^{*+}$                  &4057.7    &4107.5 &4080.79\\
 &$\frac{1}{2}$ &$D_s^-D^{*0}/D_s^+D^{*-}$        &3966.8    &4009.1 &3975.46\\
 &              &$D_s^{*-}D^{0}/D_s^{*+}D^{-}$    &3973.3    &4014.1 &3977.14\\
&0,1            &$\bar{D}^{*0}D^{*+}$             &4000.4    &4034.6 &4013.94\\
0,1,2&0         & $D_s^{*-}D_s^{*+}$              &4182.2    &4231.4 &4224.6 \\
&$\frac{1}{2}$  &$D_s^{*-}D^{*0}/D_s^{*+}D^{*-}$  &4091.3    &4133   &4119.27\\
\hline\hline
&0,1  &$\bar{B}^0B^+$                             &10569.4   &10555.8 &10559.06\\
0&0   &$B_s^-B_s^+$                               &10721.2   &10711.6 &10732.6\\
&$\frac{1}{2}$ &$B_s^-B^0/B_s^+B^-$               &10645.3   &10633.7 &10645.83\\
\hline
&0,1 &$\bar{B}^0B^{*+}$                           &10609     &10596.7 &10604.63\\
1&0  &$B_s^-B_s^{*+}$                             &10764.2   &10756   &10779.1\\
&$\frac{1}{2}$ &$B_s^-B^{*0}/B_s^+B^{*-}$         &10684.9   &10674.6 &10691.4\\
 & &$B_s^{*-}B^{0}/B_s^{*+}B^{-}$                 &10688.3   &10678.4 &10692.33\\
&0,1 &$\bar{B}^{*0}B^{*+}$                        &10648.6   &10637.6 &10650.2\\
0,1,2&0 & $B_s^{*-}B_s^{*+}$                      &10807.2   &10801   &10737.9\\
&$\frac{1}{2}$ &$B_s^{*-}B^{*0}/B_s^{*+}B^{*-}$   &10727.9   &10719.3 &10825.6
\end{tabular}}
\end{table}
\end{ruledtabular}
\end{center}

The mass spectra of the $[\bar{Q}q][\bar{q}Q]$ system are also obtained by solving the Schr\"{o}dinger
equation with Rayleigh-Ritz variational principle with the wavefunction shown in Eq.(\ref{twave}).
In spatial wavefunctions of Eq.(\ref{gem1})-(\ref{gem3}), we take the number of gaussians
$\alpha=12, ~n=7,~N=7$, and the ranges of $s_n$ for $\boldsymbol{\rho}$ are from 0.1 to 6~fm, and
0.1 to 2~fm for $\mathbf{R}$ and $\mathbf{r}$, respectively, since energy of four-quark configuration $[\bar{Q}q][\bar{q}Q]$) are converged by them, which have been discussed in detail in Ref.\cite{prd80114023}.

The results for the color-singlet channel ($\cal{D}(\cal{B})$ and $\cal{\bar{D}}(\cal{\bar{B}})$ in color
singlet) and channel coupling between color-singlet and color-octet channel are given in Table \ref{PPdb},
\ref{PVdb} and \ref{VVdb}. Since we solve the Schr\"{o}dinger equation in the finite space, so all of
them are bound states. According to the binding energy defined in Eq.(\ref{deltaE}), if it is larger
than 0, then the state should be unbound, in fact the energy of four-quark state will approaches the
threshold when the space is increased. The states with $\Delta E >0$ are marked as "$\times$" in the tables.

For the S-wave $P\bar{P}$ configuration, the quantum numbers $J^{P}$ are $0^+$. Apart from scalar meson
$\sigma$, the pseudoscalar mesons e.g. $\pi,K,\eta$ make no contribution to the $[\bar{Q}q]-[\bar{q}Q]$
systems for the parity conservation. From the TABLE \ref{PPdb}, one can easily find that the $\sigma$
meson-exchange induces attraction for the $[\bar{c}q][\bar{q}c]$ configuration but neither enough to form
stable bound state in color-singlet channel nor in color coupled channel, which marked $"1\otimes1"$ and
"coupling" in the tables. Due to the $\sigma$ meson-exchange is allowed in $s\bar{s}$, $u\bar{s}$,
$\bar{u}s$, $\bar{d}s$, or $d\bar{s}$ in CQM I, weakly bound states of $\mathcal{B}\bar{\mathcal{B}}$
system with $I=0(s)$, and 1/2 are obtained. Clearly, more bound states in $[\bar{b}q][\bar{q}b]$
configuration are found, if we take into account the effect of channel coupling. Why do the bound states of $\mathcal{B}\bar{\mathcal{B}}$
system with $I=0(s)$ appear in this case?  The spin matrix element $\langle \boldsymbol{\sigma}_{1}\cdot\boldsymbol{\sigma}_{2}\rangle=\langle \boldsymbol{\sigma}_{3}\cdot \boldsymbol{\sigma}_{4}\rangle=-3$ and others are zero. The nonzero color matrix element between color-singlet and color-octet channels  $\langle \boldsymbol{\lambda}_{1}\cdot\boldsymbol{\lambda}_{3}\rangle=\langle \boldsymbol{\lambda}_{2}\cdot \boldsymbol{\lambda}_{4}\rangle=\sqrt{32/9}$ and $\langle \boldsymbol{\lambda}_{1}\cdot\boldsymbol{\lambda}_{4}\rangle=\langle \boldsymbol{\lambda}_{2}\cdot \boldsymbol{\lambda}_{3}\rangle=-\sqrt{32/9}$. So, only coulomb  and confinement interaction, and  pairs of (1,3), (2,4), (1,4), (2,3) make contribution to the mass of $\bar{Q}Q\bar{q}q$ system. From the TABLE \ref{Distance} calculated in the CQM I, one can easily find that   $\sqrt{\langle r_{13}^2\rangle}=\sqrt{\langle r_{24}^2\rangle}=\sqrt{\langle r_{14}^2\rangle}=\sqrt{\langle r_{23}^2\rangle}$ for the $\mathcal{D}\bar{\mathcal{D}}$ system, the transition matrix
element of the coulomb  and confinement interaction canceled each other, respectively. However, for the $\mathcal{B}\bar{\mathcal{B}}$ system, the transition matrix element can't canceled each other for the different distance between quark(antiquark)-quark(antiquark). So the mass of each $\mathcal{B}\bar{\mathcal{B}}$ system
is depressed by it.
\begin{center}
\begin{ruledtabular}
\begin{table}[htp]
\caption{The binding energy of $\mathcal{D}\bar{\mathcal{D}}$ and $\mathcal{B}\bar{\mathcal{B}}$ states. The "$1\otimes1$" and "coupling" represent $\mathcal{D}\bar{\mathcal{D}}$ and $\mathcal{B}\bar{\mathcal{B}}$ in color-singlet, and coupling of color-singlet and -octet channel, respectively (unit: MeV).  The symbol "$\times$" stands for this configuration is unbound state.\label{PPdb} }
{\begin{tabular}{@{}lcccccccc@{}}
& \multicolumn{4}{c}{$\mathcal{D}\bar{\mathcal{D}}$} &
  \multicolumn{4}{c}{$\mathcal{B}\bar{\mathcal{B}}$} \\
Isospin & \multicolumn{2}{c}{CQM II} & \multicolumn{2}{c}{CQM I} &
          \multicolumn{2}{c}{CQM II} & \multicolumn{2}{c}{CQM I} \\
     & 1$\otimes$1 & coupling & 1$\otimes$1 & coupling & 1$\otimes$1 & coupling & 1$\otimes$1 & coupling\\
 \hline
I=$\frac{1}{2}$& $\times$   & $\times$  & $\times$   & $\times$   & $\times$   &$-17.5$   &$-0.2$    &$-2.6$ \\
I=1            & $\times$   & $\times$  & $\times$   & $\times$   & $\times$   &$-72.6$   &$\times$  &$-29.9$ \\
I=0 (\textit{l})& $\times$   & $\times$  & $\times$   & $\times$   & $\times$   &$-72.6$   &$\times$  &$-29.9$ \\
I=0 (\textit{s})& $\times$   & $\times$  & $\times$   & $\times$   & $\times$   &$\times$  &$-0.7$    &$-1.5$
\end{tabular}}
\end{table}
\end{ruledtabular}
\end{center}

\begin{center}
\begin{ruledtabular}
\begin{table}[htp]
\caption{The distance between each quark pairs of $\mathcal{B}\bar{\mathcal{B}}$ and $\mathcal{D}\bar{\mathcal{D}}$ in the CQM I (unit: fm).\label{Distance}}
{\begin{tabular}{@{}cccccccc@{}}
&$\sqrt{\langle r_{12}^2\rangle}$ &$\sqrt{\langle r_{34}^2\rangle}$ &$\sqrt{\langle r_{13}^2\rangle}$ &$\sqrt{\langle r_{24}^2\rangle}$ &$\sqrt{\langle r_{14}^2\rangle}$ &$\sqrt{\langle r_{23}^2\rangle}$ \\ \hline
 $\mathcal{B}\bar{\mathcal{B}}$ &0.8&0.8 &0.8 &0.8 &0.4 &1.1\\
$\mathcal{D}\bar{\mathcal{D}}$ &0.6&0.6 &6.3 &6.3 &6.3 &6.4
\end{tabular}}
\end{table}
\end{ruledtabular}
\end{center}

For the $S$-wave $P\bar{V}$ system, the $\sigma,\pi,\eta$ mesons can all be exchanged in such systems. No bound state is found in the $[c\bar{q}][\bar{c}q]^{*}$ system, if we only take into account the two-body interactions. However, the stable bound states of $[b\bar{q}][\bar{b}q]$ with $I=1,0$ are obtain in two constituent quark models. It is because that the mass of the $b$ quark is much larger than the $c$ quark, the total kinetic energy of the former is smaller than the latter, and so the meson-exchange potential can bind them to form bound states. It is reasonable to interpret the $Z_b^\pm(10610)$ and $Z_b^0(10610)$,
reported by the Belle collaboration, as the molecular state $B\bar{B}^*$. In the CQM I, in addition to the above two states, the $I=1/2$ and $0(s)$
also form two stable bound states for the $\sigma$-exchange which contributes to these channels. We also get deeply bound states for the $P\bar{V}$ system if the coupling of color-singlet and color-octet channel are taken into account.
\begin{center}
\begin{ruledtabular}
\begin{table}[htp]
\caption{The binding energy of
$\mathcal{D}^*\bar{\mathcal{D}}$ and
$\mathcal{B}^*\bar{\mathcal{B}}$ states (unit: MeV). \label{PVdb}}
{\begin{tabular}{@{}lcccccccc@{}}
& \multicolumn{4}{c}{$\mathcal{D^*}\bar{\mathcal{D}}$} &
  \multicolumn{4}{c}{$\mathcal{B^*}\bar{\mathcal{B}}$} \\
Isospin & \multicolumn{2}{c}{CQM II} & \multicolumn{2}{c}{CQM I} &
          \multicolumn{2}{c}{CQM II} & \multicolumn{2}{c}{CQM I} \\
     & 1$\otimes$1 & coupling & 1$\otimes$1 & coupling & 1$\otimes$1 & coupling & 1$\otimes$1 & coupling\\
 \hline
I=$\frac{1}{2}$& $\times$   & $\times$    & $\times$   & $\times$  & $\times$   &$-89.8$   &$-0.2$ &$-40.9$ \\
I=1            & $\times$   & $\times$    & $\times$   & $\times$  & $-1.3$     &$-164.3$  &$-1.1$ &$-107.4$ \\
I=0(\textit{l})& $\times$   & $\times$    & $\times$   & $\times$  & $-12.1$    &$-122.6$  &$-12.1$ &$-69.8$ \\
I=0(\textit{s})& $\times$   & $\times$    & $\times$   & $\times$  & $\times$   &$-47.3$   &$-1.2$ &$-3.9$
\end{tabular}}
\end{table}
\end{ruledtabular}
\end{center}

The S-wave $VV$ systems have quantum
numbers $J^{PC}=0^{++},~1^{+-}$, and $2^{++}$ for the neutral
states. The $\pi,~\eta,~\sigma$ mesons can all be exchanged between $[Q\bar{q}]^*$ and $[\bar{Q}q]^*$ for
isospin I=0(\textit{l}) and I=1, while the $\pi$-exchange is attractive just for
$I(J^{PC})=1(0^{++}),~1(1^{+-}),~0(2^{++})$. From TABLE \ref{VVdb}, one can
find that the $[c\bar{q}]^*[\bar{c}q]^*$ with quantum number $I(J^{P})=1(0^{+})$ is a stable bound state, that energy is 3996.2 MeV and 4005.7 MeV in CQM II and CQM I, respectively, if the coupling of color-singlet channel and color-octet channel
is considered. It is a possible candidate for $Z_c^{\pm}(4025)$ reported by
BESIII in the process $e^+e^-\rightarrow(D^{*}D^{*})^{\pm}\pi^{\pm}$ at a center-of-mass energy of 4.26GeV \cite{Zc4025BESIII}.
In the case of bottomonium scenario, the $[b\bar{q}]^*[\bar{b}q]^*$ with quantum numbers $I(J^{PC})=1(0^{+}),~1(1^{+}),~0(2^{+})$ are bound in color-singlet channel for the $\pi$-exchange between the
light quark pairs. The assignment of the observed state $Z_b(10650)$ to molecular state
$\mathcal{B}^*\bar{\mathcal{B}}^*$ with
$I(J^{PC})=1(1^{+})$, which has binding energy about $-1$ MeV in two quark models, is favored.
We also investigate the effect of channel coupling between
color-singlet and -octet, which are presented in TABLE \ref{VVdb},
and more deep bound states are obtained in two quark models.
\begin{center}
\begin{ruledtabular}
\begin{table}[htp]
\caption{The binding energy of
$\mathcal{D}^*\bar{\mathcal{D}}^*$ and
$\mathcal{B}^*\bar{\mathcal{B}}^*$ states (unit: MeV). \label{VVdb}}
{\begin{tabular}{@{}llcccccccc@{}}
 & & \multicolumn{4}{c}{$\mathcal{D^*}\bar{\mathcal{D}}^*$} &
  \multicolumn{4}{c}{$\mathcal{B^*}\bar{\mathcal{B}}^*$} \\
 & & \multicolumn{2}{c}{CQM II} & \multicolumn{2}{c}{CQM I} &
          \multicolumn{2}{c}{CQM II} & \multicolumn{2}{c}{CQM I} \\
 $J^P$ & $I$ & 1$\otimes$1 & coupling & 1$\otimes$1 & coupling & 1$\otimes$1 & coupling & 1$\otimes$1 & coupling\\
 \hline
&I=$\frac{1}{2}$ & $\times$ & $\times$ & $\times$  & $\times$   & $\times$   &$-172.9$  &$-0.5$    &$-108.4$\\
$0^{+}$  &I=1    & $\times$ & $-4.2$   & $\times$  & $-28.9$    & $-3.4$     & $-266.2$ &$-3.0$    &$-195.5$\\
&I=0(\textit{l}) & $\times$ & $\times$ & $\times$  & $\times$   & $\times$   &$-175.9$  &$\times$  &$-115.3$\\
&I=0(\textit{s}) & $\times$ & $\times$ & $\times$  & $\times$   & $\times$   &$-112.2$  &$\times$  &$-37.7$\\ \hline
&I=$\frac{1}{2}$ & $\times$ & $\times$ & $\times$  & $\times$   & $\times$   &$-129.0$  &$-0.3$    &$-73.9$\\
$1^{+}$&I=1      & $\times$ & $\times$ & $\times$  & $\times$   & $-1.2$     &$-201.5$  &$-0.9$    &$-144.6$\\
&I=0(\textit{l}) & $\times$ & $\times$ & $\times$  & $\times$   & $\times$   &$-160.1$  &$\times$  &$-107.6$\\
&I=0(\textit{s}) & $\times$ & $\times$ & $\times$  & $\times$   & $\times$   &$-86.0 $  &$-0.2$    &$-21.2$ \\ \hline
&I=$\frac{1}{2}$ & $\times$ & $\times$ & $\times$  & $\times$   & $\times$   &$-56.4 $  &$\times$  &$-19.3$\\
$2^{+}$&I=1      & $\times$ & $\times$ & $\times$  & $\times$   & $\times$   &$-97.0$   &$\times$  &$-62.6$\\
&I=0(\textit{l}) & $\times$ & $\times$ & $\times$  & $\times$   & $-11.1$    &$-133.0$  &$-11.0$   &$-95.1$\\
&I=0(\textit{s}) & $\times$ & $\times$ & $\times$  & $\times$   & $\times$   &$-41.0$   &$-1.1$    &$-5.7$
\end{tabular}}
\end{table}
\end{ruledtabular}
\end{center}

About eighteen $XYZ$ charmonium-like resonances discovered by Belle, BaBar, BES and LHCb collaborations, and many resonances have been confirmed by different experiments. A well established one among them is the $X(3872)$, which was first discovered in 2003 by
Belle Collaboration \cite{x3872prl91262001} in the $\pi^+\pi^-J/\psi$ invariant mass spectrum in the process
$B\rightarrow K\pi^+\pi^-J/\psi$, and later confirmed by six other experiments \cite{x3872prl93072001,x3872prl93162002,x3872epjc721972,x3872prd71071103,x3872jhep04154,x3872prl112092001}.
Its quantum number have been studied by Belle, BaBar, CDF and LHCb, and determined to be
$I^GJ^{PC}=0^{+}1^{++}$ \cite{x3872prl110222001}. However, from the discussions above it is clear that few bound state exists in $c\bar{q}\bar{c}q$ system either for CQM I or for CQM II, even if the channel coupling of color-singlet and color-octet is taking into account. In order to obtain bound state of $c\bar{q}\bar{c}q$ system, we believe that the four-quark interactions that would not be factorable as a sum of two-body potentials should be included, for example,
the multi-body confinement in the four-quark Hamiltonian \cite{prd82074001,prd90054009,prd92034027}, or enlarged
the Hilbert space of the four-quark problem, include the compact diquark-antidiquark ($\delta-\bar{\delta}$) pair \cite{prl91232003,prd71014028,prd89114010,PRL113112001,prd95074007,prd94094041} and so on.

\section{Summary}\label{summary}
Using two chiral quark models, we dynamically study the mass spectra of the $[\bar{Q}q][\bar{q}Q]$ system
by a high accuracy numerical method based on Gaussian expansion method.
No stable bound state is obtained in
$[\bar{c}q][\bar{q}c]$ if we only take color-singlet channel into
account. However, since the mass of $b$ quark is heavier than $c$
quark, and so the total kinetic energy of the former is smaller than the
latter, the meson-exchange can provide enough attraction to bind $[b\bar{q}]$ and
$[\bar{b}q]$ for $B^*\bar{B}^*$ with $I(J^{PC})=1(0^{++}),~1(1^{+-}),~0(2^{++})$
and $B\bar{B}^*$ with isospin $I=1$, or $0$. The $Z_b^{\pm}(10610)$, $Z_b^{0}(10610)$ and
$Z_b^{\pm}(10650)$ reported by Belle collaboration
can be reasonably interpreted as a molecular state $B\bar{B}^*$ and $B^*\bar{B}^*$,
respectively. In CQM I, $\sigma$-exchange allowed between  $u$, $d$ and $s$ quark,
can provide enough attraction to form stable bound states $[b\bar{s}][s\bar{b}]$,
and $[b\bar{q}][s\bar{b}]$ for the case of $P\bar{P}$, $P\bar{V}$ and
$V\bar{V}$. The channel coupling of color-singlet and color-octet are also
discussed above, we find more deep bound states in
$\mathcal{B}\mathcal{\bar{B}}$, $\mathcal{B}\mathcal{\bar{B}}^*$
and $\mathcal{B}^*\mathcal{\bar{B}}^*$ system. A bound state
$[c\bar{q}]^*[q\bar{c}]^*$ with $I(J^{PC})=1(0^{++})$ is especially obtained in our
calculation, which is a possible candidate for $Z_c^{\pm}(4025)$ \cite{Zc4025BESIII} reported by
BESIII. However, we can not find a candidate for each $X(3872)$, $Z_c^{\pm}(3900)$
and other $XYZ$ charmonium-like states in our calculation. To obtain the bound states of $c\bar{c}\bar{q}q$ system, we believe that the four-quark interactions that would not be factorable as a sum of two-body potentials should be
included, or the Hilbert space should be enlarged.

\acknowledgments{This work is supported partly by the National Science Foundation of China (under Contracts Nos.11265017,
11675080, 11775118, 11535005, 11475085, and 11690030), and the China Postdoctoral Science Foundation (under Grant No.2015M571727), and by the Guizhou province outstanding youth science and technology talent cultivation object special funds (under Grant No. QKHRZ(2013)28).}


\end{document}